\documentstyle[preprint,tighten,epsfig,aps]{revtex}
\begin{document}
\baselineskip 1.1 \baselineskip

\def\beqar{\begin{eqnarray}}
\def\ra{\rightarrow}
\def\eeqar{\end{eqnarray}}
\def\be{\begin{eqnarray}}
\def\ee{\end{eqnarray}}
\def\beqast{\begin{eqnarray*}}
\def\eeqast{\end{eqnarray*}}
\def\be{\begin{enumerate}}
\def\ee{\end{enumerate}}
\def\lag{\langle}
\def\rag{\rangle}
\def\fnote#1#2{\begingroup\def\thefootnote{#1}\footnote{#2}
\addtocounter{footnote}{-1}\endgroup}
\def\beq{\begin{equation}}
\def\eeq{\end{equation}}
\def\haf{\frac{1}{2}}
\def\pa{\partial}
\def\plb#1#2#3#4{#1, Phys. Lett. {\bf #2B}, #3 (#4)}
\def\npb#1#2#3#4{#1, Nucl. Phys. {\bf B#2}, #3 (#4)}
\def\prd#1#2#3#4{#1, Phys. Rev. {\bf D#2}, #3 (#4)}
\def\prl#1#2#3#4{#1, Phys. Rev. Lett. {\bf #2}, #3 (#4)}
\def\mpl#1#2#3#4{#1, Mod. Phys. Lett. {\bf A#2}, #3 (#4)}
\def\rep#1#2#3#4{#1, Phys. Rep. {\bf #2}, #3 (#4)}
\def\llp#1#2{\lambda_{#1}\lambda'_{#2}}
\def\lplp#1#2{\lambda'_{#1}\lambda'_{#2}}
\def\slash#1{#1\!\!\!\!\!/}
\def\rpv{\slash{R_p}~}
\def\hs{\hat s}
\def\ca{{\cal A}}
\def\cb{{\cal B}}
\def\cc{{\cal C}}


\draft
\preprint{
\begin{tabular}{r}
KAIST-TH 97/21
\\
hep-ph/yymmddd
\end{tabular}
}
\title{
$B \ra X_s {l_i}^+ {l_j}^-$ Decays with R-parity Violation
}
\author{
Ji-Ho Jang
\thanks{E-mail: jhjang@chep6.kaist.ac.kr},
Yeong Gyun Kim
\thanks{E-mail: ygkim@chep6.kaist.ac.kr}
and
Jae Sik Lee
\thanks{E-mail: jslee@chep6.kaist.ac.kr}
}
\address{
Department of Physics, Korea Advanced Institute of Science and
Technology \\
Taejon 305-701, Korea \\
}

\maketitle

\begin{abstract}
We derive the upper bounds on certain products of R-parity- and
lepton-flavor-violating couplings from
$B \ra X_s {l_i}^+ {l_j}^-$ decays.
These modes of $B$-meson decays
can constrain the product combinations of the couplings with
one or more heavy generation indices
which are comparable with or stronger than the present bounds.
From the studies of the invariant dilepton mass spectrum and
the forward backward asymmetry of the emitted leptons
we note the
possibility of detecting R-parity-violating signals even when
the total decay rate due to R-parity violating couplings 
is comparable with that in the standard model and
discriminating two types of R-parity-violating signals. 
The general expectation of the enhancement
of the forward backward asymmetry of the emitted leptons
in the minimal supersymmetric standard model 
with R-parity may be corrupted by R-parity violation.
\end{abstract}


{\bf 1.}
In supersymmetric extensions of the standard model, there are
gauge invariant interactions which violate 
the baryon number ($B$) and the lepton number ($L$)
in general. To prevent occurrings of these $B$- and $L$-violating
interactions in supersymmetric extensions of the standard model,
the additional global symmetry is required.
This requirement leads to the consideration of
the so called R-parity($R_p$).
Even though the requirement of $R_p$ conservation makes a theory
consistent with
present experimental searches, there is no good theoretical justification
for this requirement. Therefore
the models with explicit $R_p$-violation have been considered by
many authors \cite{ago}.

To discover the $R_p$ violation in future experiments,
we need to know what kinds of
couplings are severely constrained by present experimental data.
Therefore it is important to constrain the $R_p$ violating
couplings from the present data, especially data on the processes forbidden
or highly suppressed in the SM.
Usually the bounds on the $R_p$ violating couplings
with at least two heavy fields
are not stronger than those with at most one heavy field.

In this paper, we derive
the upper bounds on certain products of $R_p$ and
lepton flavor violating couplings from $b \ra s {l_i}^+ {l_j}^-$ decays
in the minimal supersymmetric
standard model (MSSM) with explicit $R_p$ violation.
These modes of $B$-meson decays
can constrain the product combinations of the couplings with
one or more heavy generation indices.

In the MSSM,
the most general $R_p$-violating superpotential is given by
\beq
W_{R\!\!\!\!/_p}=\lambda_{ijk}L_iL_jE_k^c+\lambda_{ijk}'L_iQ_jD_k^c+
\lambda_{ijk}''U_i^cD_j^cD_k^c.
\eeq
Here $i,j,k$ are generation indices and we assume that possible bilinear terms
$\mu_i L_i H_2$ can be rotated away.
$L_i$ and $Q_i$ are the $SU(2)$-doublet lepton and quark superfields and
$E_i^c,U_i^c,D_i^c$ are the singlet superfields respectively.
$\lambda_{ijk}$ and
$\lambda_{ijk}''$ are antisymmetric under the interchange of the first two and
the last two generation indices respectively; $\lambda_{ijk}=-\lambda_{jik}$ and
$\lambda_{ijk}''=-\lambda_{ikj}''$. So the number of couplings is 45 (9 of the
$\lambda$ type, 27 of the $\lambda'$ type and 9 of the $\lambda''$ type).
Among these 45 couplings, 36 couplings are related with the lepton
flavor violation.

There are upper bounds on a {\it single} $R_p$ violating coupling from
several different sources \cite{han,beta,numass,agagra}.
Among these, upper bounds from
neutrinoless double beta decay \cite{beta}, $\nu$ mass \cite{numass} and
$K^+,t-$quark decays \cite{agagra} are strong.
Neutrinoless double beta decay gives
$\lambda'_{111}<3.5\times10^{-4}$.
The bounds from $\nu$ mass are $\lambda_{133}<3\times10^{-3}$
and $\lambda'_{133}<7\times10^{-4}$.
From $K^+$-meson decays one obtain $\lambda'_{ijk}<0.012$ for $j=1$ and 2.
These bounds from $K^+$-meson decays are basis-dependent \cite{agagra,bha}.
Here all masses of scalar partners which mediate the processes are assumed
to be 100 GeV.
Extensive reviews of the updated limits on a single $R_p$-violating
coupling can be found in \cite{bha,chahu}.

There are more stringent bounds on some products of
the $R_p$-violating couplings from the mixings of the
neutral $K$- and $B$- mesons and rare leptonic decays of
the $K_L$-meson, the muon and the tau \cite{choroy}, $B^0$ decays
into two charged leptons \cite{Lee}, $b\bar{b}$ productions at LEP \cite{Feng}
and muon(ium) conversion, and $\tau$ and $\pi^0$ decays \cite{Ko}.

In this paper, we assume that the baryon number 
violating couplings $\lambda''$'s
vanish in order to avoid too fast proton decays.  Especially in the models with
a very light gravitino ($G$) or axino ($\tilde{a}$), $\lambda''$ have to be
very small independently of $\lambda'$ 
from the proton decay $p\rightarrow K^+G~({\rm or}~K^+\tilde{a})$ 
; $\lambda''_{112}<10^{-15}$ \cite{Choi}. 
One can construct a grand unified model which has only lepton number
non-conserving trilinear operators in the low energy superpotential when
$R_p$ is broken only by bilinear terms of the form $L_i H_2$
\cite{hasu}. And usually it may be very difficult to discern signals of
$B$-violating interactions above QCD backgrounds \cite{numass}.

\vspace{1.0cm}
{\bf 2.}
The exchange of the sleptons or squarks leads to the four-fermion
interactions in the effective lagrangian.
Among these four-fermion operators, there are
terms relevant for $b \ra s {l_i}^+ {l_j}^-$.
These effective terms have 2 down-type quarks and 2 charged leptons.
From Eq. (1) we obtain
\beqar
{\cal L}^{eff}_{\rpv}&=&
\sum_{n=1}^{3}\frac{2}{m_{\tilde{\l}_n}^2}\left[\lambda^*_{njk}\lambda'_{nlm}
(\bar{e}_jP_Re_k)(\bar{d}_mP_Ld_l)
+h.c.\right]
\nonumber \\
&-& \sum_{n,r,s=1}^{3}\frac{1}{2m_{\tilde{Q}_n}^2}K_{nr}K^*_{ns}
\lambda'^*_{jrk}\lambda'_{lsm}
(\bar{e}_j\gamma^{\mu}P_Le_l)(\bar{d}_m\gamma_{\mu}P_Rd_k),
\eeqar
where $K$ is the CKM matrix and we assume
the matrices of the soft mass terms are diagonal and the CKM matrix comes from
the mixings between down type quarks.
From this Lagrangian, one finds that 
the various semileptonic b decays could appeare at tree level;
$ b \ra q e^+ e^-, q \mu^+ \mu^-, q \tau^+ \tau^-, q e^{\pm} \mu^{\mp},
q \mu^{\pm} \tau^{\mp}, q e^{\pm} \tau^{\mp} ~(q = d, s)$.

At presents, the measurements of the branching ratios 
of the $b \ra s {l_i}^+ {l_j}^-$ 
processes give the upper bounds (at 90 $\%$ C.L.)\cite{cleo}
\beqar
BR( b \rightarrow s e^+ e^-) &<& 5.7 \times 10^{-5}, \nonumber \\
BR( b \rightarrow s \mu^+ \mu^-) &<& 5.8 \times 10^{-5}, \nonumber \\
BR( b \rightarrow s e^{\pm} \mu^{\mp}) &<& 2.2 \times 10^{-5}.
\eeqar
In the SM, the process $b \rightarrow s e^{\pm} \mu^{\mp}$ 
is forbidden due to the conservation of each lepton flavor number.
On the other hand, 
the decay $b \ra s \mu^+ \mu^-$ ($b  \ra s e^+ e^-$)
is dominated by the electroweak penguin
and receives small contributions from box diagrams and magnetic penguins.
A recent analysis gives
$BR(b \rightarrow s e^+ e^-)_{SM} = (8.4 \pm 2.3) \times 10^{-6}$,
$BR(b \rightarrow s \mu^+ \mu^-)_{SM} = (5.7 \pm 1.2) \times 10^{-6}$
\cite{ali}.
The experimental bounds are almost one order of magnitude larger than 
the standard model expetations.
If we neglect the SM contribution, 
the decay rate of the processes $b \ra s {l_i}^+ {l_j}^-$ reads
\beqar
\Gamma(b \ra s {l_i}^+ {l_j}^-) =
\frac{m_b^5}{6144 \pi^3 \tilde{m}^4} \left[
\vphantom{\frac{m_b^5}{6144 \pi^3 \tilde{m}^4}} 
4(|{\cal A}_{ij}|^2 + |{\cal B}_{ij}|^2)
+ |{\cal C}_{ij}|^2 \right]
\eeqar 
The constants ${\cal A}_{ij},{\cal B}_{ij}$ and ${\cal C}_{ij}$
are given by
\beqar
{\cal A}_{ij}&=& \sum_{n=1}^{3}\lambda^{*}_{nji}\lambda'_{n32}, \nonumber \\
{\cal B}_{ij}&=& \sum_{n=1}^{3}\lambda_{nij}\lambda'^{*}_{n23}, \nonumber \\
{\cal C}_{ij}&=& \sum_{n,p,s=1}^{3}K_{np}K_{ns}^*
\lambda'^{*}_{jp3}\lambda'_{is2} =
\sum_{n=1}^{3}
\lambda'^{*}_{jn3}\lambda'_{in2} .
\eeqar
Note that
we assume the universal soft mass $\tilde m$ and ignore the lepton mass.
To remove the large uncertainty in the total decay rate associated with 
the $m_b^5$ factor, it is convenient to normalize 
$BR(b \ra s {l_i}^+ {l_j}^-)$ to the semileptonic rate
$BR(b \ra c e \bar{\nu})$.
We then obtain
\beqar
4(|{\cal A}_{ij}|^2 + |{\cal B}_{ij}|^2) + |{\cal C}_{ij}|^2 =
\frac{6144 {\tilde m}^4 {G_F}^2 {|V_{cb}|}^2 f_{\rm ps}(m_c^2/m_b^2)}{192}
\frac{BR(b \ra s {l_i}^+ {l_j}^-)}{BR(b \ra c e \bar{\nu})},
\eeqar
where $f_{\rm ps}(x) = 1 - 8 x + 8 x^3 - x^4 - 12 x^2~ \ln x$  is 
the usual phase-space factor. 
Inserting into Eq. (6) the semileptonic rate 
$BR(b \ra c e \bar{\nu}) \approx 10.5 \%$, 
$f_{\rm ps}(m_c^2/m_b^2) \approx 0.5$,
$|V_{cb}| \approx 0.04$,
we obtain 
\beqar
4(|{\cal A}_{ij}|^2 + |{\cal B}_{ij}|^2) + |{\cal C}_{ij}|^2 =
3.3 \times 10^{-3} ~(\frac{\tilde m}{100~ GeV})^4~ Br(b \ra s {l_i}^+
{l_j}^-).
\eeqar

In the case of $b \ra s e^+ e^-$ decay ($i=1, j=1$), we obtain
\beqar
4(|{\cal A}_{11}|^2 + |{\cal B}_{11}|^2) + |{\cal C}_{11}|^2 <
1.9 \times 10^{-7},
\eeqar
from the Eq. (7) and the upper limit on the branching ratio, Eq. (3). 
Under the assumption that only one product combination is not zero,
we can get the upper bounds 
on the following combinations of the $\lambda  {\lambda}^{\prime}$-
and ${\lambda}^{\prime} {\lambda}^{\prime}$-type; 
\beqar
|\lambda_{n11} \lambda'_{n32}| < 2.2 \times 10^{-4}, \nonumber\\  
|\lambda_{n11} \lambda'_{n23}| < 2.2 \times 10^{-4}, \nonumber\\ 
|\lambda'_{1n3} \lambda'_{1n2}| 
< 4.3 \times 10^{-4}.
\eeqar
For the product combinations of $\lambda \lambda'$ type, 
we observe that the bounds on 
$\lambda_{121} \lambda'_{232}$,
$\lambda_{131} \lambda'_{332}$,
$\lambda_{121} \lambda'_{223}$,  
$\lambda_{131} \lambda'_{323}$
are stronger than the previous bounds (see Table I).

In a similar way, another upper bounds can be obtained in the case of 
$b \ra s \mu^+ \mu^-$ and $b \ra s e^{\pm} \mu^{\mp}$ decays.
For the $b \ra s \mu^+ \mu^-$ decay, the bound on  $\lplp{233}{232}$ is 
stronger
than the previous bounds. And for the $b \ra s e^{\pm} \mu^{\mp}$ decay,
the bounds on $\lplp{233}{232}$ and all product combinations of $\llp{}{}$ 
type are stronger than the previous bounds.

In fact, the $B_s \ra {l_i}^+ {l_j}^-$ process is described with
the same parameters $\cal{A}$,$\cal{B}$,$\cal{C}$  
and in some cases gives slightly more stingent bounds \cite{Lee}.
But, two things make the decay $b \ra s l^+_i l^-_j$ more useful than
the $B_s \ra {l_i}^+ {l_j}^-$ process.
One thing is that 
the contribution of the product combinations of $\lambda' \lambda'$ type
is vanishing in the limit of zero lepton mass 
and so these parameters cannot be constrained 
in the case of the $B_s \ra {l_i}^+ {l_j}^-$ process.
The other thing is that the experimental upper limit on 
the branching ratios exist only in the $B_s \ra \mu^+ \mu^-$ process
at present.  

\vspace{1.0cm}
{\bf 3.}
The structure of $\rpv$ effective lagrangian (Eq. (2)) is
quite different from that of the SM, it would be interesting to compare the
invariant dilepton mass spectrum and the forward backward asymmetry of the
emitted leptons in the presence of $\rpv$ with
those of the SM \cite{ali,SMbsll} 
and the minimal MSSM with $R_p$ \cite{MSSMbsll}. The forward backward
asymmetry is defined by
\beqar
\frac{d\ca (\hs)}{d\hs}=\int_0^1 \frac{d^2\cb}{d\hs dz}
-\int_{-1}^0 \frac{d^2\cb}{d\hs dz}, \nonumber 
\eeqar
where $\hs\equiv s/m_b^2$,
$s$ is the invariant mass of the lepton pair,
and $z\equiv \cos\theta$ is the angle of $l^+$ measured
with respect to the $b$-quark direction in the dilepton center of mass system.

Neglecting the masses of the
strange quark and leptons, we obtain the $\rpv$ invariant dilepton mass
spectrum as follows
\beq
\frac{d\cb^{\rpv}(b\ra s l_i^+ l_j^-)}{d\hs}=\frac{ {\cb}_{sl}(1-\hs)^2
\left[24(|\ca_{ij}|^2+|\cb_{ij}|^2)\hs+|\cc_{ij}|^2(1+2\hs)\right] }
{16 f_{\rm ps}(m_c^2/m_b^2) G_F^2 {\tilde m}^4 |V_{cb}|^2},
\eeq
where $\cb_{sl}$ is the semileptonic rate $BR(b\ra ce{\bar \nu})$.
We find that there are {\it no} interferences between contributions of the SM and
the $R_p$-violating model under consideration. So, the total invariant dilepton
mass spectrum is given by the direct sum of the SM contributions and
$R_p$-violating contributions.
In Fig. 1 (a), we show the normalized invariant dilepton mass spectrum
$\frac{d\overline{\cb}}{d\hs}\equiv\frac{1}{\cb}\frac{d\cb}{d\hs}$.
The solid line denotes the typical SM prediction \cite{ali,SMbsll} 
and dashed and
dotted lines the $R_p$-violating contributions to the invariant dilepton mass
spectrum. 
Since the structure of the effective lagrangian
corresponding to $\llp{}{}$-type $R_p$ violation
is different from that of $\lplp{}{}$-type one, the $\hs$ dependences of 
the invariant dilepton mass spectrum due to these
two types of $R_p$ violation are quite different from each other. The $\hs$
dependence of $\lplp{}{}$-type $R_p$ violation is nearly the same as that of
the SM and this type of $R_p$ violation
enhances the decay rate on the low $\hs$ region. But the
behavior of the contribution from
$\llp{}{}$-type $R_p$ violation is different from the SM and
this type of $R_p$ violation
enhances the invarian dilepton mass spectrum
in the region around $\hs=1/3$. Therefore, from
the invariant dilepton mass spectrum 
one can detect the $R_p$ violating signals even if
the magnitude of $R_p$ violating coupling is comparable with that of the SM and
it is possible to discriminate two types of $R_p$ violating signals, $\llp{}{}$-
and $\lplp{}{}$-type.

We also obtain $\rpv$ 
forward backward asymmetry of the emitted leptons
\beq
\frac{d\ca^{\rpv}(b\ra s l_i^+ l_j^-)}{d\hs}=\frac 
{-3 \cb_{sl} |\cc_{ij}|^2 (1-\hs)^2 \hs}
{32 f_{\rm ps}(m_c^2/m_b^2) G_F^2 {\tilde m}^4 |V_{cb}|^2}.
\eeq
Let us note that sign of this asymmmetry is negative. 
In Fig. 1 (b), we show the normalized forward backward 
asymmetry of the emitted leptons
$\frac{d\overline{\ca}}{d\hs}\equiv\frac{1}{\cb}\frac{d\ca}{d\hs}$.
There is no contribution to the asymmetry from the $\llp{}{}$-type $R_p$ 
violation. But there is {\it negative} contribution from $\lplp{}{}$-type $R_p$ 
violation which can compensate the SM asymmetry. 
Since there are no interferences between contributions of the SM and
the $R_p$-violating model under consideration,
the total asymmetry $\frac{d\tilde{\ca}}{d\hs}$ is given by
\beq
\left(\frac{d\tilde{\ca}}{d\hs}\right)\equiv
\frac{\left(\frac{d{\ca}}{d\hs}\right)^{\rm SM}+
      \left(\frac{d{\ca}}{d\hs}\right)^{\rpv}}
     {\left(\frac{d{\cb}}{d\hs}\right)^{\rm SM}+
      \left(\frac{d{\cb}}{d\hs}\right)^{\rpv}}=
\frac{\left(\frac{d\overline{\ca}}{d\hs}\right)^{\rm SM}+
\xi_{\llp{}{}}\left(\frac{d\overline{\ca}}{d\hs}\right)^{\rpv}_{\llp{}{}}+
\xi_{\lplp{}{}} \left(\frac{d\overline{\ca}}{d\hs}\right)^{\rpv}_{\lplp{}{}}}
     {\left(\frac{d\overline{\cb}}{d\hs}\right)^{\rm SM}+
\xi_{\llp{}{}}\left(\frac{d\overline{\cb}}{d\hs}\right)^{\rpv}_{\llp{}{}}+
\xi_{\lplp{}{}}\left(\frac{d\overline{\cb}}{d\hs}\right)^{\rpv}_{\lplp{}{}}},
\eeq
where $\xi_{(\llp{}{},\lplp{}{})} 
\equiv \cb^{\rpv}_{(\llp{}{},\lplp{}{})}/\cb^{\rm SM}$ and
$\left(\frac{d\overline{\ca}}{d\hs}\right)^{\rpv}_{\llp{}{}}=0$. 
In the case $\xi_{\llp{}{}} \gg \xi_{\lplp{}{}}$, the SM asymmetry will be
diluted by the factor of $\sim1/(1+\xi_{\llp{}{}})$.
In the case $\xi_{\llp{}{}} \ll \xi_{\lplp{}{}}$, even the sign of the asymmetry
could be different from the prediction of the SM depending on the size
of $\xi_{\lplp{}{}}$.  For illustration, we show 
the above forward backward
asymmetry with $\xi_{\llp{}{}}=1$ or $\xi_{\lplp{}{}}=1$ in Fig. 2.
So, also from the forward backward asymmerty
of the emitted leptons one can detect the $R_p$ violating signals even if
the magnitude of $R_p$ violating coupling is comparable with that of the SM and
it is possible to discriminate two types of $R_p$ violating signals, $\llp{}{}$-
and $\lplp{}{}$-type.

From the studies of the
process $b\ra s l^+ l^-$ in the minimal MSSM with $R_p$ \cite{MSSMbsll},
the SUSY signal may 
appear as the enhancement of this asymmetry by more than 100 \% relative 
to SM expectations.
But, in the presence of $R_p$ violation this
minimal MSSM enhancement may be corrupted
\footnote{In fact, the precise definition of the asymmetry
in the first reference of \cite{MSSMbsll} is different from ours. But, our
conclusion is not changed by this difference.}.

\vspace{1.0cm}
{\bf 4.}
To conclude, we have derived the more strigent 
upper bounds on certain products of $R_p$- and lepton-flavor-violating 
couplings from the recent measurements of $b \ra s {l_i}^+ {l_j}^-$ decay
rates at CLEO. 
From the studies of the invariant dilepton mass spectrum and the forward
backward asymmetry of the emitted leptons, we note the
possibility of detecting $R_p$-violating signals even when
the total decay rate due to $R_p$-violating couplings 
is comparable with that in the SM and
discriminating two types of $R_p$-violating signals, 
$\llp{}{}$- and $\lplp{}{}$-type. The general expectation of the enhancement
of the forward backward asymmetry of the emitted leptons
in the minimal MSSM with $R_p$ may be
corrupted by R-parity violation.

\section*{acknowledgements}
We thank P. Ko and K. Choi for helpful discussions.
This work was supported in part by the KRF postdoctorial program
(Y.G.K.) and in part by the KAIST Basic Science Research Program (J.S.L.).

\begin{table}
\caption{\label{haha}
Upper bounds on the magnitudes of products of couplings
derived from $b \rightarrow s l_i^+ l_j^-$.
}
\begin{tabular}{llll}
Decay Mode & Combinations Constrained & Upper bound & Previous bound \\
\hline
$b \rightarrow s e^+ e^-$&
$\llp{121}{232}$ & 2.2$\times 10^{-4}$ & 8.0 $\times 10^{-3~a}$
\\
&$\llp{131}{332}$ &2.2$\times10^{-4}$  & 2.9 $\times 10^{-2}$
\\
&$\llp{121}{223}$ &2.2$\times10^{-4}$  & 6.0 $\times 10^{-4~b}$,
~~(9.0 $\times 10^{-3~c}$)
\\
&$\llp{131}{323}$ &2.2$\times10^{-4}$  & 7.2 $\times 10^{-4~b}$,
~~(1.2 $\times 10^{-2~c}$)
\\
&$\lplp{113}{112}$ &4.3$\times10^{-4}$  & 1.4 $\times 10^{-4~b}$,
~~(4.0 $\times 10^{-4~c}$)
\\
&$\lplp{123}{122}$ &4.3$\times10^{-4}$  & 1.4 $\times 10^{-4~b}$,
~~(4.0 $\times 10^{-3~c}$)
\\
&$\lplp{133}{132}$ &4.3$\times10^{-4}$  & 1.1 $\times 10^{-4~a}$
\\
\hline
$b \rightarrow s \mu^+\mu^-$&
$\llp{122}{132}$ & 2.2$\times 10^{-4}$ & 5.5 $\times 10^{-5~d}$
\\
&$\llp{232}{332}$ &2.2$\times10^{-4}$  & 5.5 $\times 10^{-5~d}$
\\
&$\llp{122}{123}$ &2.2$\times10^{-4}$  & 5.5 $\times 10^{-5~d}$
\\
&$\llp{232}{323}$ &2.2$\times10^{-4}$  & 5.5 $\times 10^{-5~d}$
\\
&$\lplp{213}{212}$ &4.4$\times10^{-4}$  & 1.4 $\times 10^{-4~b}$,
~~(8.1 $\times 10^{-3~c}$)
\\
&$\lplp{223}{222}$ &4.4$\times10^{-4}$  & 1.4 $\times 10^{-4~b}$,
~~(3.2 $\times 10^{-2~c}$)
\\
&$\lplp{233}{232}$ &4.4$\times10^{-4}$  & 2.5 $\times 10^{-2~a}$
\\
\hline
$b \rightarrow s e^{\pm} \mu^{\mp}$&
$\llp{122}{232}$ & 1.4$\times 10^{-4}$ & 1.8 $\times 10^{-2~a}$
\\
&$\llp{132}{332}$ &1.4$\times10^{-4}$  & 2.9 $\times 10^{-2}$
\\
&$\llp{121}{132}$ &1.4$\times10^{-4}$  & 8.0 $\times 10^{-3~a}$
\\
&$\llp{231}{332}$ &1.4$\times10^{-4}$  & 2.9 $\times 10^{-2}$
\\
&$\llp{122}{223}$ & 1.4$\times 10^{-4}$ & 6.0 $\times 10^{-4~b}$,
~~(9.0 $\times 10^{-3~c}$)
\\
&$\llp{132}{323}$ &1.4$\times10^{-4}$  & 7.2 $\times 10^{-4~b}$,
~~(1.2 $\times 10^{-2~c}$)
\\
&$\llp{121}{123}$ &1.4$\times10^{-4}$  & 6.0 $\times 10^{-4~b}$,
~~(1.0 $\times 10^{-2~c}$)
\\
&$\llp{231}{323}$ &1.4$\times10^{-4}$  & 7.2 $\times 10^{-4~b}$,
~~(1.2 $\times 10^{-3~c}$)
\\
&$\lplp{113}{212}$ &2.7$\times10^{-4}$  & 1.4 $\times 10^{-4~b}$,
~~(1.8 $\times 10^{-3~c}$)
\\
&$\lplp{123}{222}$ &2.7$\times10^{-4}$  & 1.4 $\times 10^{-4~b}$,
~~(3.6 $\times 10^{-2~c}$)
\\
&$\lplp{133}{232}$ &2.7$\times10^{-4}$  & 1.1 $\times 10^{-4~a}$
\\
&$\lplp{213}{112}$ &2.7$\times10^{-4}$  & 1.4 $\times 10^{-4~b}$,
~~(1.8 $\times 10^{-3~c}$)
\\
&$\lplp{223}{122}$ &2.7$\times10^{-4}$  & 1.4 $\times 10^{-4~b}$,
~~(3.6 $\times 10^{-3~c}$)
\\
&$\lplp{233}{132}$ &2.7$\times10^{-4}$  & 2.5 $\times 10^{-2~a}$
\\
\end{tabular}
a: Bounds from $B \rightarrow X_c~ l \bar{\nu}$\cite{jang}. 
b: Considering the bounds from
    $K \rightarrow \pi \nu {\bar \nu}$\cite{agagra}. 
c: Ignoring the bounds from
    $K \rightarrow \pi \nu {\bar \nu}$. 
d:  Bounds from $B_s \rightarrow \mu^+
    \mu^-$\cite{Lee}.
Others : See the second reference of \cite{bha}.
\end{table}

\newpage
\begin{figure}[ht]
\hspace*{-1.0 truein}
\psfig{figure=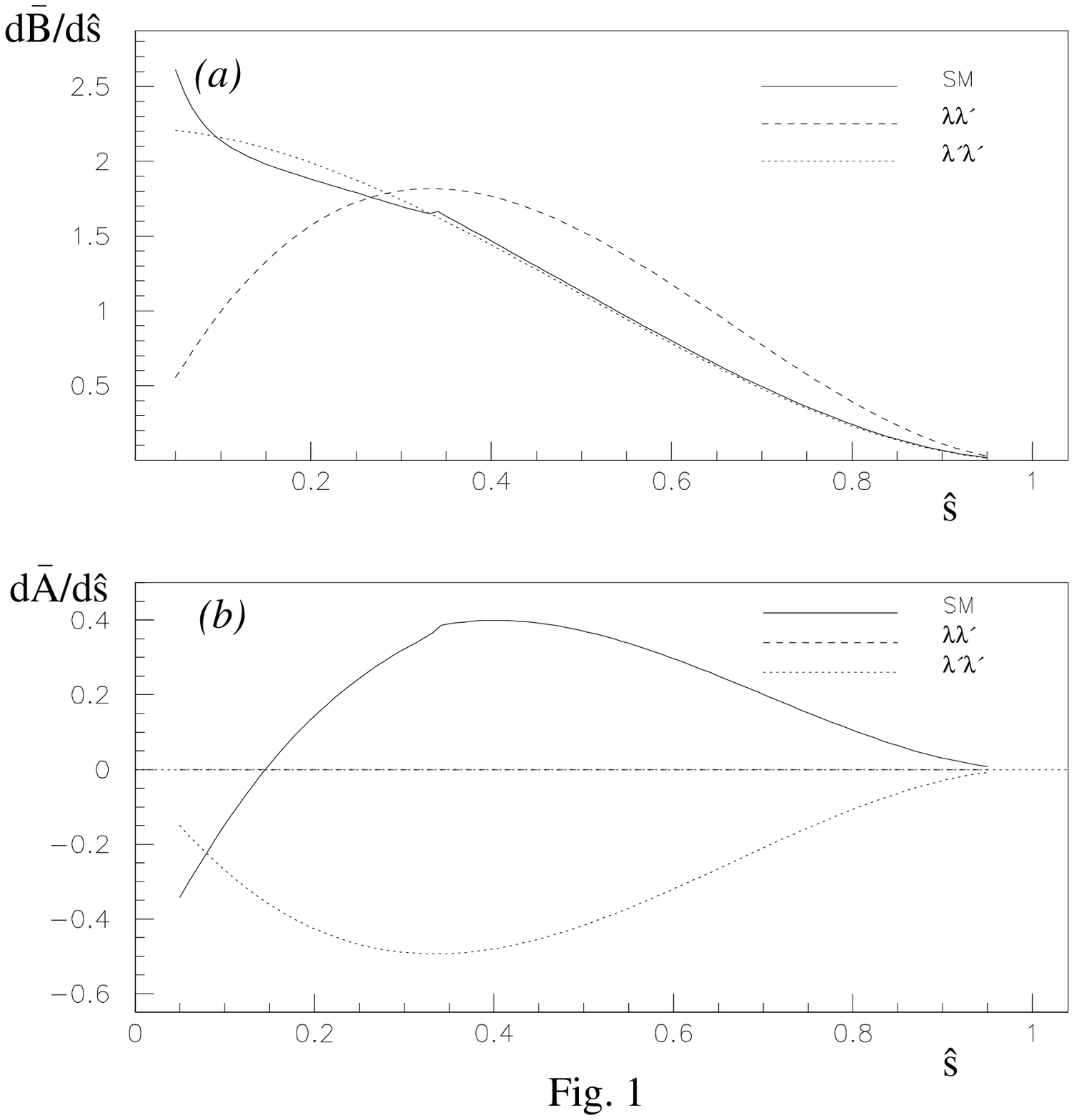}
\caption{
Plots of normalized invariant dilepton mass spectrum (a) and 
forward backward asymmetry of the emitted leptons (b).
The solid line denotes the typical SM prediction [16,18] and dashed and
dotted lines the $R_p$-violating contributions, Eq.(10) and Eq. (11).
}
\label{fig1}
\end{figure}

\newpage
\begin{figure}[ht]
\hspace*{-1.0 truein}
\psfig{figure=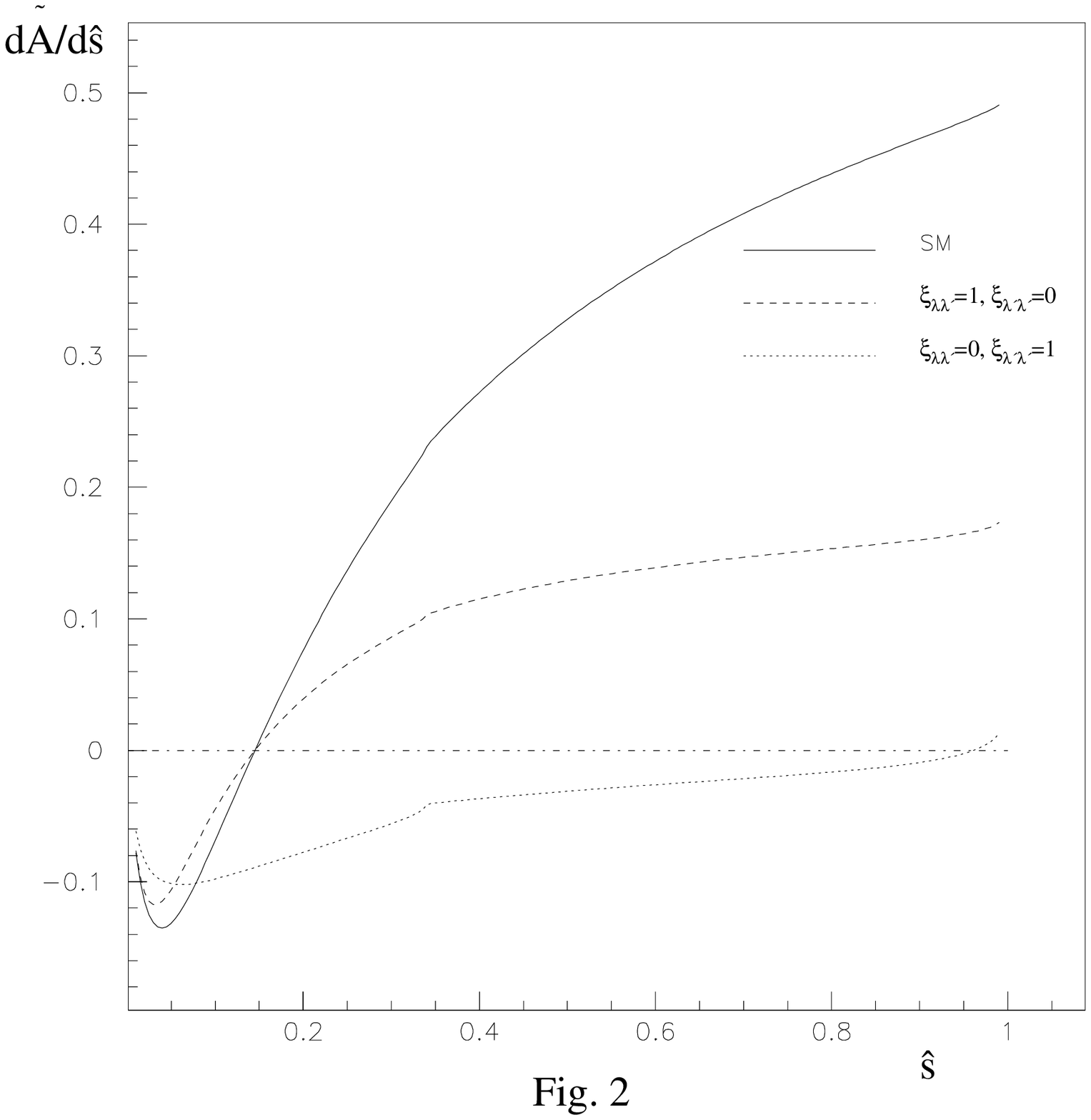}
\caption{
Plots of total forward backward asymmetry of the emitted leptons, 
$\left(\frac{d\tilde{\ca}}{d\hs}\right)$.
The solid line denotes the typical SM prediction [16,18] and dashed and
dotted lines the $R_p$-violating contributions, Eq.(12).
}
\label{fig2}
\end{figure}

\end{document}